# "Meaning" as a sociological concept:
# A review of the modeling, mapping, and simulation of the communication of knowledge and meaning

Loet Leydesdorff
Amsterdam School of Communications Research (ASCoR), University of Amsterdam
Kloveniersburgwal 48, 1012 CX Amsterdam, The Netherlands;
loet@leydesdorff.net; http://www.leydesdorff.net

**Abstract**

The development of discursive knowledge presumes the communication of meaning as analytically different from the communication of information. Knowledge can then be considered as a meaning which makes a difference. Whereas the communication of information is studied in the information sciences and scientometrics, the communication of meaning has been central to Luhmann's attempts to make the theory of *autopoiesis* relevant for sociology. Analytical techniques such as semantic maps and the simulation of anticipatory systems enable us to operationalize the distinctions which Luhmann proposed as relevant to the elaboration of Husserl's "horizons of meaning" in empirical research: interactions among communications, the organization of meaning in instantiations, and the self-organization of interhuman communication in terms of symbolically generalized media such as truth, love, and power. Horizons of meaning, however, remain uncertain orders of expectations, and one should caution against reification from the meta-biological perspective of systems theory.

**Keywords**: semantic map, anticipation, self-organization, communication, meaning

**Introduction**

The communication of meaning as different from the communication of information can perhaps be considered as the *differentia specifica* of social systems. Whereas biological systems are sometimes able to provide meaning to information and thus shape a semantic domain (Maturana, 1978; Maturana & Varela, 1980), and human minds can reflexively change the meaning of information, the capacity to communicate both information and meaning can be considered as an evolutionary achievement in inter-human languaging (Luhmann, 2002). When meaning can be communicated, this communication can further be codified and discursive knowledge also developed.

The issue is beset with conceptual difficulties. First, the distinction between information and meaning can be conflated by defining information in terms insufficiently independent from meaning: as "meaningful information" or, in Bateson's (1972: 489) formulation, as "a difference which makes a difference." Shannon-type information is defined as a series of differences (in a probability distribution), whereas such differences can only make a next-order difference for a receiving (or observing) system that is able to provide the information with meaning. Secondly, "meaning" in the sociological tradition is often considered primarily as a subjective category and is not sufficiently understood in terms of inter-subjective communication.

Third, the distinction between the communication of meaning and the communication of knowledge needs further elaboration. I shall argue that knowledge can be considered as "a meaning which makes a difference," whereas meaning is generated when first-order differences (Shannon-type information) make a difference for a receiving system. However, the communication of knowledge requires the relative closure of the discourse in terms of specific codes of communication.

These issues and distinctions are particularly salient nowadays, given the emergence of a knowledge-based economy (Foray, 2004; Leydesdorff, 2006a). Unlike a political economy, which can be considered as based on interactions among (*i*) economic exchange relations and (*ii*) political arrangements, the additional dynamics of (*iii*) knowledge-based communication requires the structural organization of the sciences (Whitley, 1984; Dasgupta & David, 1984). A third structured subdynamics can then be added to inter-human communication at the level of society. This third subdynamics potentially disturbs the relative stabilization of political economies in nations and tends to globalize and meta-stabilize existing market relations (Leydesdorff & Zawdie, 2010).

**The distinction between information and meaning**

Shannon (1948, at p. 3) provided a lucid distinction of meaning from information at the beginning of his paper entitled *A Mathematical Theory of Communication*:

> Frequently the messages have *meaning*; that is, they refer to or are correlated according to some system with certain physical or conceptual entities. These semantic aspects of communication are irrelevant to the engineering problem. The

> significant aspect is that the actual message is one *selected from a set* of possible messages.

Two systems of reference are distinguished in this quotation: the formal one of the electrical engineer or, in other words, Shannon himself as a mathematician, and the possible meanings provided by substantive discourses. Shannon considers the latter as irrelevant for the definition of information, but he notes the significance of potentially different *selection mechanisms*.

Shannon-type information is defined as *yet* content-free (Theil, 1972). This condition of "still-to-be-provided-with-meaning" by a system of reference is also manifested in the units of measurement (e.g., bits of information) which are dimensionless. Shannon-type information does not yet contain meaning other than the *mathematical* definition of the expected information value contained as uncertainty in a message as a finite series of differences—in other words, a probability distribution. What the expected information content of the distribution means can only be defined by an observing system using its own selection mechanism. Meaning is defined "in use" (Wittgenstein, 1953). Note that selection by an observing system is deterministic and system-specific, whereas variation can be stochastic.

The meaning provided to the (Shannon-type) information can sometimes reduce uncertainty. Reduction of uncertainty can be measured as negentropy (Brillouin, 1962): this possibility originates from the difference which the difference (or a series of differences, that is, a probability distribution) can make for a receiving system. Thus, "a difference which makes a difference" (Bateson, 1972) can also reduce the uncertainty that prevails and be identified as meaningful or observed information, given the specification of a system of reference.

Weaver (1949, at p. 116) noted that Shannon's abstract definition of information as uncertainty might sound "bizarre," but that this level of abstraction might also be needed to develop a theory of meaning. Meaning is generated in use by specific systems that are able to receive and/or process meaning. This receiving system, however, does not have to be an observer, but can also be a discourse. Information is then provided with meaning which may contain a supra-individual and coordinating function. In other words, the meaning is codified *among* human beings, that is, at the intersubjective level.

The systems-theoretical tradition has focused on the observer as the individual unit of analysis (e.g., Edelman, 1989). For example, Von Foerster (1979) ascribed to Maturana as his "Theorem Number One" that "Anything said is said *by* an observer," and added his own corollary that "Anything said is said *to* an observer." He concluded that the two observers share a language. From this perspective, however, language is considered only as a meta-biological domain.

Maturana (1978, at p. 49), for example, noted that for an observer a "second-order consensual domain […] becomes indistinghuishable from a semantic domain." However, understanding in language (as both an observer and a participant; Giddens, 1979) was set aside by him as the language of a human "super-observer" (pp. 58f.). From this biological perspective, languaging—linguistic behavior—can be observed. The dynamics of human language as different from communication among insects cannot properly be analyzed from these meta-biological perspectives (Epstein and Axtell, 1996).

Human language—and more generally, communication—connects not only observers, but also their observational reports, that is, the *translation of their observations into communication* provides the messaging with intersubjective meaning, and this codification allows for the communication of subjective meaning at the supra-individual level (Mead, 1934; Pask, 1975). Distinguishing the observational reports in language from the observers making these "utterances"[1] moves us from the realm of mathematical biology and psychology into the sociological realm of communication as interhuman, that is, meaningful and potentially knowledge-based coordination.

**Language and symbolic mediation**

Meaning is generated in a system when different pieces of information are related as messages to one another, for example, as words in sentences (Hesse, 1988; Law & Lodge, 1984). The information is then positioned in a network with an emerging (and continuously reconstructed) structure. This positioning can be done by an individual who—as a system of reference—can provide personal meaning to the events, but meaning can also be provided at the supra-individual level, for example, in a discourse. In the latter case, meaning is discursive, and its dynamics can therefore be expected to be different from those of psychological meaning.

Whereas a psychological identity can be expected to strive to integrate a plurality of meanings that could be provided to single events (for example, in order to avoid "cognitive dissonance"), the social system—as a *dividuum* (Fuchs, 1998, at pp. 225 ff.; cf. Latour, 2002; Nietzsche, 1878: I, §57)—can tolerate the entertainment of different meanings, and has the additional option to differentiate itself into subsystems which codify these meanings differently. This plurality in rationalities can be functional to the processing of complexity in a pluriform society (cf. Boudon, 1979; Bourdieu, 2004).

For example, politicians and economists can discuss "shortages of energy" although among physicists "energy" is considered as a *conserved* quantity. Codifications facilitate and speed up the communication by making the communication system-specific. At the market, for example, one can simpy pay the price of something without having to negotiate. The price codifies the value of the commodity. Prices make it possible to abstract from the underlying values in another semantic domain (for example, that of banking).

This possibility of functional differentiation in the codes of communication and the potentially symbolic generalization of meaning was first elaborated in the tradition of social-systems theory by the sociologists Talcott Parsons and Niklas Luhmann. Building on Durkheim's argument that norms function as integrative at the supra-individual level, Parsons theorized the possibility of functional differentiation among the roles of agents in different subsystems of society. According to Parsons (e.g., 1961, at p. 41), collectivities of roles can complement one another in fulfilling various functions in society. However, Parsons himself did not make a connection to what I consider as his other major contribution, namely, the theory of the symbolic generalization of the media of communication (Parsons, 1968). In his scheme, differentiation remains confined to four major meta-biologically derived functions (adaptation, goal-attainment, integration, and latency).

---

[1] Luhmann ([1984], 1995) defined communication in terms of three elements: (1) utterance, (2) information, and (3) understanding. Such a definition, in my opinion, does not distinguish sufficiently between communication among animals (e.g., insects) and human beings (Leydesdorff, 2006b). Luhmann's theory can be considered as unnecessarily non- or even anti-humanistic (Leydesdorff, 2000, 2010c).

Following Merton (1957), Luhmann ([1984], 1995) historicized social functions and proposed that they develop as specific rationalities in the different and historically variable processings of meaning in social subsystems. Luhmann added the evolutionary perspective that new forms of codification can be invented; for example, at first coins were used, then banknotes, and much later credit cards. Each communication subsystem develops further by overwriting and repositioning the previous versions of its coding.

The prime example of this cultural evolution of communication has been provided by Kuhn's (1962) notion of paradigm shifts. "Phlogiston," for example, was backgrounded in scholarly discourse once "oxygen" was constructed as a new concept (Priestly, 1774-1777). The new paradigm (in chemistry) opened domains for puzzle-solving and further communication with a code that is different from the previous one. However, both "oxygen" and "phlogiston" use a code of communication very different from exchange processes on the market, which obey economic mechanisms of exchange. Analogously, the truth of a scientific statement is different from the religious truth of a dogma. Other dimensions of interhuman communication (e.g., affection, power) always also play a role, although the institutional setting may facilitate the functionality of specific codes among these symbolically generalized media more than others. For example, it is transgressive to favour one scientific theory over another for political reasons, or to bribe a judge.

Interhuman communication can thus be considered as a fabric woven in many directions: each communication can be provided with meaning in terms of power (Foucault, 1966), economic utility, affection, scientific truth, etc. These latent dimensions of communication resound and operate selectively in all interhuman communication. However, in specific communications some selections can be expected to operate more strongly than others because of the functionality of coding. Symbolically generalized codes enable us to be specific in our communications and thus to process more complexity.

The selecting codes of communication are not a given, but enacted and reconstructed in use as the culturally and therefore supra-individually constructed dimensions of communication. Historically, the gradual disintegration of the Holy Roman Empire at the end of the Middle Ages, first in the Investiture Contest and then during the Reformation, generated an additional degree of freedom in which the prevailing form of differentiation among the codes was changed from a hierarchical (stratified) one to another in which the codes could be reconstructed more freely as a function of the communication.

The tight coupling between institutions and functions could then gradually be loosened to the extent that institutions could also be reorganized in terms of their functions. For example, during the Scientific Revolution of the 17$^{th}$ century and after the trial of Galileo, scientific truth could be differentiated from religious truth. Analogously, the *trias politica*—developed during the 18$^{th}$ century (Montesquieu, 1748)—regulates the differentiation among political discourses in modern constitutions. This functional differentiation in the communication has been institutionalized in national constitutions since the American and French Revolutions.

Herbert Simon (1973) hypothesized that any evolving system can be expected to operate with an alphabet. Thus, one might hypothesize 20+ symbolically generalized media of communication possible in interhuman communications. These codes of communication should not be reified: they are historically constructed and enacted bottom-up in interhuman communications, but as they are reconstructed recursively over time, they can be expected to function as control mechanisms at the level of society that enable us to enrich our communication by allowing for greater precision. Note that the "top" or next-order level is not yet defined or fixed in terms of these bottom-up (re)construction processes.

The codes operate as selection mechanisms by enabling us to focus the communication. Selection mechanisms can reinforce one another in processes of mutual shaping (McLuhan, 1964). Thus, selection mechanisms can be expected to shape historical trajectories that are relatively stable (for example, in institutions). A next-order selection may drive a local stabilization into global meta-stabilization, or into regimes which function with dynamics that differ from—since they counteract as feedback mechanisms on—the dynamics of historical developments. Such further differentiation among selection mechanisms (stabilization and globalization) can uncouple the communication reflexively from the historical process in which it emerges.

For example, communication with money first speeds up the communication on the market to the extent that local forms of capitalism can be shaped. Bank notes, stock exchanges, and credit cards provide means for worldwide transactions with correspondingly increasing speeds and precision. Marx already identified this emerging mechanism in capitalism as "alienating." Luhmann ([1984], 1995) proposed to study the dynamics of communication (cf. Marx's "exchange value") as analytically different from the dynamics of human or group behavior ("use value"). The systems of reference are altered by the change of perspective caused by the newly emerging code of communication. This potential globalization is an attribute of the communication and not of the communicators.

In the terminology of *autopoiesis* theory (Maturana & Varela, 1980), the two dynamics of processing meaning—at the level of agency and at the social level—remain "structurally coupled" and "interpenetrate" each other reflexively (Luhmann, 2002). However, one can expect the cybernetics of communication to be different from the dynamics of human (group) behavior. Communications, for example, can travel worldwide without the communicators as carriers having to move.

**Horizons of meaning**

Interhuman communication is based on interactions among both communicators and communications. The meanings interact in a non-linear dynamics which is not hardwired and therefore no longer necessarily subject to the second law of thermodynamics. The redundancies generated by the processing of meaning, for example, can structure and reconstruct the information processing from the perspective of hindsight. The possibility of such non-linear dynamics is enabled by language as an evolutionary achievement: meaning can proliferate discursively at a speed much faster than its instantiations in language (e.g., in fantasies and wishes; cf. Weinstein and Platt, 1969) because of the possible feed-forward loops between individual experiences and expectations, and communication in language.

Whereas the biological *autopoiesis* processes the history of the communications in terms of their structural sediments—for example, in terms of differentiations among organs or species—the orientation toward horizons of possible communication is provided by the additional communication of meaning in language. The linguistically or symbolically mediated communication channels are changed by the historical communication in terms of the communications *possible* thereafter. In other words, redundancies—sets of possibilities—are generated. Unlike the biological *autopoiesis* of the living (Maturana & Varela, 1980), meaning can be communicated reflexively and with reference to and in anticipation of "horizons of meaning." I use the plural of this Husserlian concept in order to emphasize that one can expect the horizons of meaning to be structured by symbolically generalized codes of communication.

Luhmann (1986) criticized Schultz's (1932; 1953) interpretation of Husserl's phenomenology because of the emphasis on observable instantiations of the cybernetics of behavior in the *life-world* (cf. Habermas, 1981; Habermas & Luhmann, 1971). Husserl himself had formulated his philosophy as a "transcendental phenomenology" with an emphasis on expectations (as different from observations). From the perspective of hindsight, this focus on intentions, meanings, and expectations can be considered as "mathematical" in the sense that it enables us to intuit other possible dimensions which resound in the empirical events that have happened to occur historically (Derrida, 1974; Heidegger, 1962; Husserl, [1935/6] 1962). We have no access to the possible other than by placing what exists reflexively between brackets. Husserl used the Greek word epochè ($\dot{\epsilon}\pi o \chi \eta$) for this "suspension" of all judgments about the existence of an external world.

The analytical specification of expectations before proceeding to actual observations enables us to specify whether observed differences (variation) can be considered as significant. This is formalized, for example, in the chi-square test of statistical significance using a theory of measurement. Husserl (1935/6) noted that the positivistic focus on observables had eroded this basis of the modern sciences, and that one should instead return to the reflexive position of Descartes, but reconstruct it in order to ground the social sciences in reflexivity.

Not incidentally, therefore, Husserl (1929) called the book in which he explained his intersubjective "horizons of meaning" *Cartesian Meditations*. Husserl's reference is to Descartes' distinction between *res extensa* and *res cogitans*. The Cartesian *Cogito* knows him/herself as uncertain and different from the external world. In this act of doubt, the contingent *Cogito* finds the transcendent environment as the Other or a personal God. However, Husserl doubted this next step: the *cogitatum* of the *Cogito* is not necessarily God, but can also be considered as an intersubjective domain to which we all have personal access: the horizons of meaning that we share (to different extents). This domain is not in the *res extensa*, but remains *res cogitans*. In other words, the meaning that we provide to the events does not "exist" physically, but incurs on us as one among a set of culturally possible meanings.

In the social sciences—as in the other theoretical sciences—one can use models to specify the expectations. The specification of expectations makes future states available in the present as potentially meaningful. The model—as different from the individual intuition—enables us to communicate about these future states with greater precision by invoking the symbolic codes of scholarly discourse. A model thus is part of discursive knowledge; it can be improved by argumentative contributions. However, this providing of new meaning is highly codified; only those who understand the model can contribute meaningfully. The model enables these participants to entertain a communication among specialists in which further knowledge can be developed and exchanged. In other words, the model is part of a communicative reality.

This communicative reality that the communicators shape over time and reflexively reconstruct cannot be considered as *res extensa*, but belongs to the *res cogitans*; it is not stable like matter, but remains in flux like language. It enables us to communicate in terms of uncertainties (e.g., possibly relevant questions) and expectations. Husserl (1929) recognized this realm as *cogitatum*, that is, the substance about which the *Cogito* remains uncertain. Our mental predicates provided to the world in intersubjective exchanges with intentional human beings, shape our culture and therewith ground what Husserl also called a "concrete ontology" or, in other words, "a universal philosophy of science" (1929, at p. 159).

This philosophy of science enables us to understand scientific models and concepts as specifically coded meanings that we attribute to an external reality at the intersubjective level. Note that from this perspective, the external world is not a social construct as in post-modernism; it is a cognitively hypothesized and highly codified construct that can be accessed reflexively and perhaps partly reconstructed by individual agency, but only in terms of further communications. In the *res extensa*, resources can be mobilized (for example, at the institutional level), but such policies can succeed only insofar as they enable us to access, deconstruct, and reconstruct the self-organization of cultural constructs in the *res cogitans*.

The models and not the modeled substance shape the sciences as cultural artifacts. According to Husserl's (1935) *The Crisis of the European Sciences*, however, an empiristic self-understanding prevails in the modern sciences. In order to move the social sciences forward, one has to stay with the transcendental that one can retrieve in one's self and thus recognize the sciences as part and parcel of a realm of cultural expectations that can be communicated. The "naturally" observed or perceived at the individual level is shaped by and rewritten in a realm of intersubjective expectations and their possible communication. Note that this *res cogitans* is also *res*, that is, real, and thus a possible subject of empirical investigation. The predictions on the basis of the models, for example, can be expected to feed back (e.g., technologically) on our material life.

**The structuration of expectations**

The scientific model as an exchange mechanism of cognitive expectations can provide us with a heuristics to understand the communication of other, for example, normative expectations. Normative exchanges can be expected to shape, for example, political discourse. In political discourse, events are provided with meanings that differ from those given be scientific discourse. In other words, social order is not a given, but a set of variously codified expectations that interact and self-organize in the *res cogitans*. This order of expectations can be sustained by institutions which function as instantiations (Giddens, 1984). Political discourse, for example, can be focused in a parliamentary debate, whereas scholarly discourse can be retrieved in scientific journals. The complex and internally differentiated order of expectations remains latent; the instantiations can be considered as their co-variations at specific moments of time.

The codes can be considered as the latent dimensions that structure the discourses in analytically different directions. Two levels can be distinguished in this structuration: the codes operate as the internal axes of meaning-providing structures. This operation is recursive: meaning is provided to the information contained in the events, and meaning can further be codified—that is, provided with symbolic meaning—in the communication. Under the condition of functional differentiation the axes can be expected to span horizons of meanings in increasingly orthogonal directions. Some codes can in a next selection be generalized symbolically or, from the perspective of their stabilization along trajectories, be globalized as horizons of meaning that feed back on the local meaning processing in interhuman communication.

Luhmann (1986) provided a thorough elaboration of Husserl's concept of "intersubjectivity" in sociological terms. Three levels were distinguished in the communication of meaning: (*i*) local interactions, (*ii*) organization of meaning in historical instances, and (*iii*) the self-organization of the codes of communication. In modern, pluriform societies self-organization in different directions can be expected to prevail—communications are no longer coordinated at the center—while in pre-modern societies communication was organized in terms of institutions. Organization integrates, while self-organization tends to differentiate the functions of the communication. Organization operates at specific moments of time—according to Luhmann (2000), by making decisions—and self-organization of meaning operates over time as codification in fluxes of communication.

The two cybernetic mechanisms of organization (integration at interfaces) and self-organization (differentiation of codes) can be considered as the woof and warp of the evolution of the *cogitatum* in a multidimensional space. This evolutionary development is driven bottom-up by variation in the interactions, while the codes operate in terms of selection mechanisms. For example, economic exchanges are organized in terms of local markets, but can self-organize a global market equilibrium if left sufficiently free to do so. Scientific communication is organized within communities and institutions, but these communities compete in hypothesizing and following the dynamics at the level of scientific fields (Bourdieu, 1976, 2004). Love and affection can be organized in terms of marriages, but also otherwise (Luhmann, 1982). Organization of meaning is historically contingent.

**The factor model and latent semantic analysis**

Functionally differentiated codes of communication cannot be observed directly, but their operations can be reconstructed reflexively as latent variables. Not incidentally, it was Paul Lazarsfeld, Merton's colleague at Columbia University, who first made the connection between the latency of communication and the factor model. A communication network can be rewritten as a matrix. The matrix representation of this network allows us to decompose the main dimensions of a communication system using factor analysis. In other words, the network is composed in terms of relations, but it contains a structure (technically, a so-called "eigenstructure") in which every relation can be positioned. Factor analysis enables us to investigate this latent structure in terms of the so-called eigenvectors of the matrix.

For example, words can be considered as meaningful information when they are placed in relation to one another in sentences. The sentences can be expected to contain meaning because information is specifically related. (Note that this can also be done by an observer, but my focus here is on language as communication.) A textual unit can be provided with specific meanings, for example, in scholarly, political or other discourses. The symbolically generalized codes operate as next-order selection mechanisms shaping, for example, paradigms. Three selection mechanisms can thus be hypothesized as operating in parallel: a first one positioning the relational information in an observing system (at each moment), a second one coding meaning in potentially different directions (over time), and a third one potentially globalizing the stabilizations of meaning in terms of different codes of communication (Lucio-Arias and Leydesdorff, 2009a).

At first, this problem may seem intractable: when both the variations and the (factorial) structures change over time, one obtains a system of partial differential equations with no easy solution. However, the problem can be decomposed. The factor model can bring latent dimensions into view. By reorganizing the data in this multi-dimensional space, one can draw a semantic map in terms of the observables (e.g., words). The organizing principles (that is, the dimensions) remain latent, but can, for example, be penciled into the map.

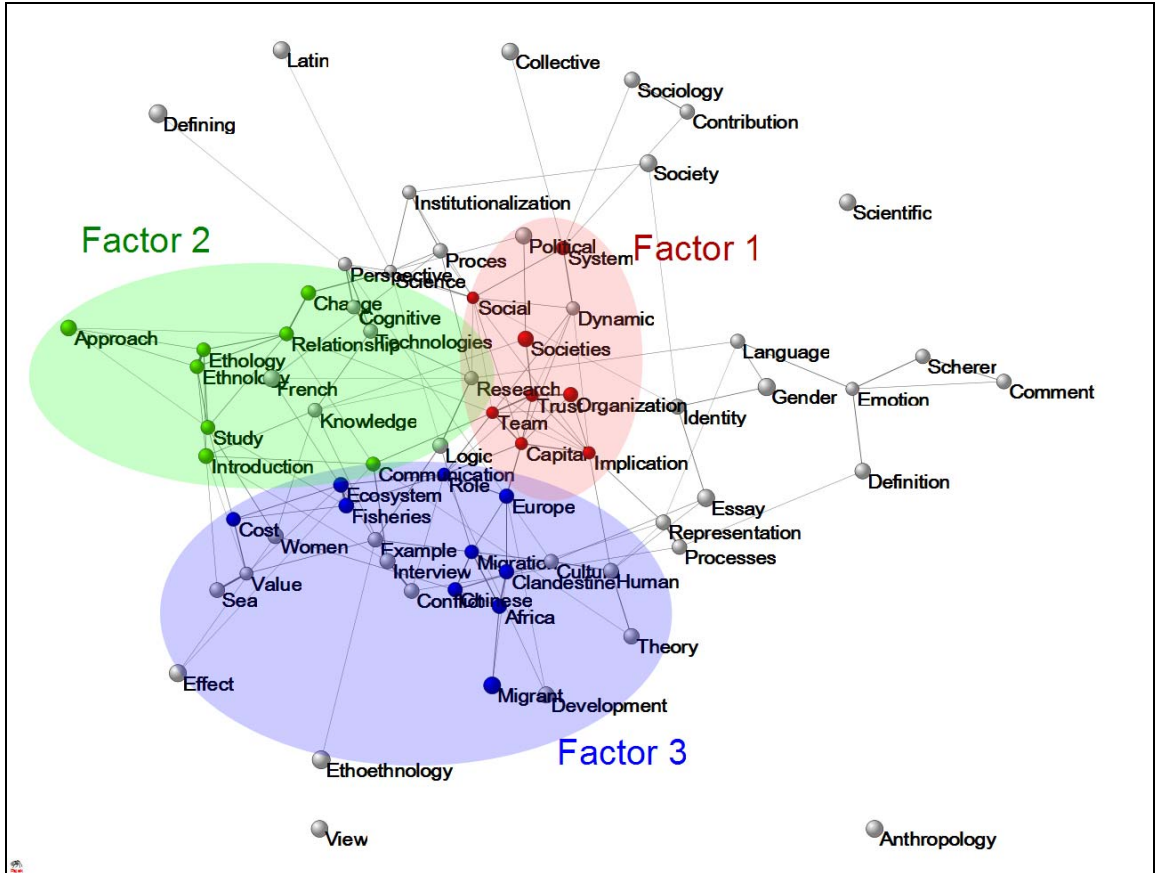

**Figure 1**: Semantic map among 56 title words connected at *cosine* ≥ 0.1 among 149 titles of documents in *Social Science Information* 2005-2009.

Figure 1, for example, shows a semantic map using the title words of 149 documents which were published in *Social Science Information* during the period 2005-2009. In these titles, 69 words occurred three or more times, of which 66 were related in the normalized word-document matrix [2] at a level of *cosine* > 0.1.[3] Factor analysis of this word-document matrix enabled me to color the main dimensions into the map as shown here for the first three factors with red, green, and blue, respectively. (These first three factors explain in this case only 10.8% of the common variance).

---

[2] The words were normalized using the observed/expected ratios instead of the raw data (Leydesdorff & Welbers, 2011). One can add the margin totals and grand sum to the word-document matrix and compute the expected value for each cell ($e_{ij}$) from the observed ones ($o_{ij}$) using $e_{ij} = \dfrac{\sum_i o_{ij} \sum_j o_{ij}}{\sum_i \sum_j o_{ij}}$. The routine for generating these semantic maps can be found at http://www.leydesdorff.net/ti/index.htm.
[3] I use the non-parametric *cosine* instead of the Pearson correlation for the representation of the multi-dimensional space (Salton & McGill, 1983; Ahlgren *et al*., 2003). All distributions are skewed because of the prevailing selection pressures.

Note that this is not a relational map as in social network or co-word analysis. The words "science" and "social," for example, co-occur most strongly by far in this domain of titles, *viz*., 12 times. However, the similarity in the distributions of occurrences among "capital" and "trust" (which co-occur only four times) is much greater than that between "science" and "society". In this dataset, the relation between "trust" and "capital" is more important—as a correlation in the semantic space of correlations—than the relation between "science" and "society." Technically, this semantic space can also be called a vector space, as against the relational space of the observable network of relations. A vector space is generated when a network of relations is spanned from which an architecture necessarily emerges.

One can take this reasoning one step further and also position the factors (or so-called "eigenvectors") in the vector space. Figure 2 shows the result of this projection for the same data. The (orthogonally rotated) factor matrix is used as input to a network visualization program. (Negative factor loadings are indicated with dotted lines.[4])

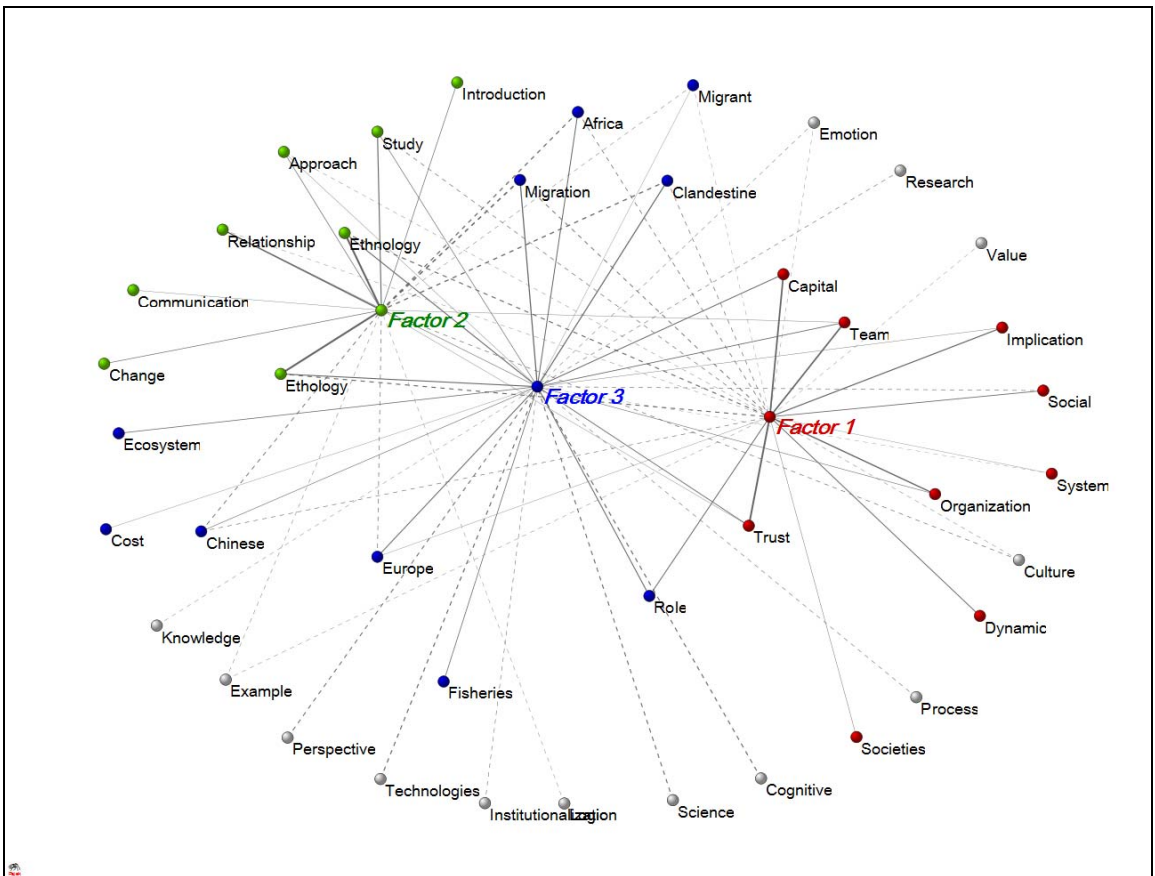

**Figure 2**: Three-factor solution of 40 words in titles of 149 documents published in *Social Science Information* 2005-2009 (words with factor loadings between -0.1 and 0.1 excluded; Fruchterman & Rheingold, 1991).

---

[4] Factor loadings between -0.1 and +0.1 were suppressed (in SPSS) in order to enhance the visualization.

Figure 2 enables us to designate more easily the three specialties involved: Factor 1 focuses on issues of organizational sociology, Factor 2 on ethnological studies, and Factor 3 on migration studies and the Mediterrenean. In other words, these clusters are reconstructed as representations of the three dominant repertoires publishing during this period in *Social Science Information.* Note that there are no positive correlations between Factor 1 and Factor 2. Between Factor 1 and Factor 3, however, the words "Role" and "Capital" provide articulation points; between Factors 2 and 3 the word "Ethology" is an articulation point between the star-formed graphs representing the discursive differentiation. The various discourses are *organized* within the context of this journal.

**The knowledge-based dynamics**

Factor analysis provides a static representation of the multi-dimensional space in an instantiation; in the case above, this was the aggregate of a five-year period. Dynamically, the next question is whether these factors of meaning *organization* are integrated into a single body of knowledge or whether differentiation is indicated. In the case of differentiation, a synergy can sometimes be found despite a lack of integration among the main dimensions. Thus, one can oppose coherence and synergy: coherence is an indicator of organization, and synergy an indicator of *self-organization*. Using this distinction, the variables (words) and structures (factors) are considered as a *system,* and thus a systemic measure is needed.

Mutual information in three or more dimensions—sometimes called "configurational information"—can be used as an indicator of the potential synergy (e.g., Abramson, 1963; Ashby, 1964; Jakulin, 2005; Leydesdorff & Sun, 2009; McGill, 1954; Ulanowicz, 1986). Although this information measure is expressed in bits, it is not a Shannon-type information (Krippendorff, 2009; Yeung, 2008, at p. 59). However, Garner & McGill (1956, at p. 227) suggested that the information measure is more useful than the interaction variance which behaves similarly, but assumes a normal distribution.

Both interaction variance and mutual information in three (or more) dimensions can be either positive or negative (Garner & McGill, 1956; Leydesdorff, 2010d). Negative values indicate a reduction of uncertainty at the systems level. This reduction cannot be expected historically (because the second law is equally valid for probabilistic entropy),[5] but has to be self-organized within the system against the axis of time. A decrease in uncertainty (that is, a negative value for this entropy) means that redundancy generated by *evolutionary* self-organization in the globally organized knowledge base is prevalent over the *historical* organization. Note that these are systems measures which cannot be attributed to individual components without using aggregation rules (Leydesdorff *et al.*, 2006; Leydesdorff & Fritsch, 2006; Theil, 1972).

The mutual information among the three factors in the system of title words studied above was +50.6 millibits. The positive value indicates that the synergy among the three factors is not found in the historical organization and integration in the titles of the journal. When the analysis, however, is repeated for the 187 documents that cite one of the 149 documents published in *Social Science Information*, I obtain Figure 3, and mutual information in three dimensions of –106.2 millibits of information.[6]

---

[5] The second law of thermodynamics holds equally for probabilistic entropy, since $S = k_B H$ and $k_B$ is a constant (the Boltzmann constant). Because of the constant, the development of $S$ over time is a function of the development of $H$, and *vice versa.*

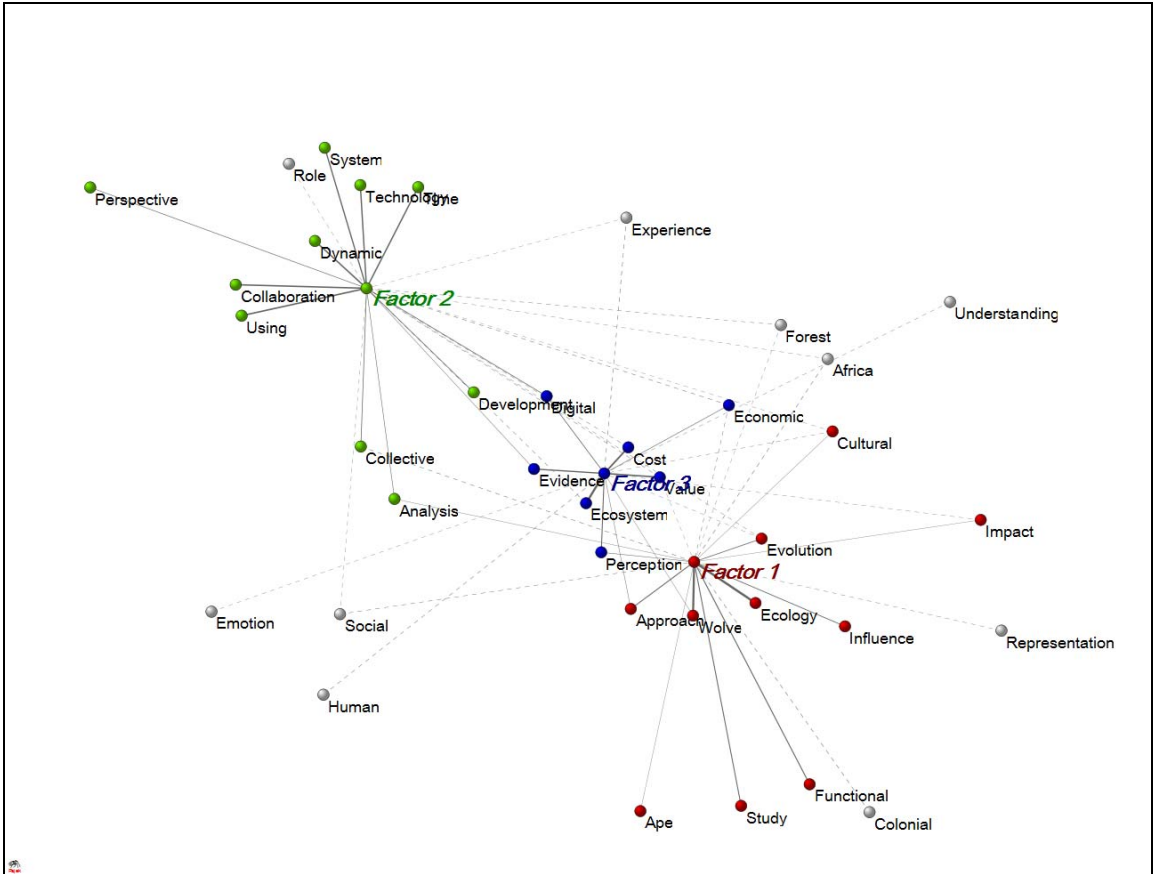

**Figure 3**: Three-factor solution of 40 words in the titles of 187 documents citing *Social Science Information* 2005-2009 (factor loadings between -0.1 and 0.1 excluded; Fruchterman & Rheingold, 1991).

The citing articles belong to different literatures: Factor 1 indicates evolutionary theorizing, Factor 2 represents studies about technology and society, and Factor 3 studies about evolving systems. "Digital" and "Evidence" function as articulation points between Factors 2 and 3, and only "Approach" between Factors 1 and 3. Although these three citing literatures are not integrated, a synergy in their differentiation is indicated by the negative value of the mutual information in three dimensions (Lucio-Arias & Leydesdorff, 2009b).

In other words, the evolutionary differentiation is reduced in the organizational integration of title words in *Social Science Information*. In addition to this organization, the self-organization can be made visible in a map as differentiation because it is instantiated. However, the differentiation develops over time and has therefore to be measured algorithmically. The dynamic development of the virtual structure can thus be indicated using these information measures.

---

[6] Krippendorff's (2009) information interaction measure $I_{ABC \rightarrow AB:AC:BC}$ was 120.0 mbits and 89.9 mbits for the original documents and the citing ones, respectively. Thus, the two sets are not so very different in this respect (Leydesdorff, 2010d).

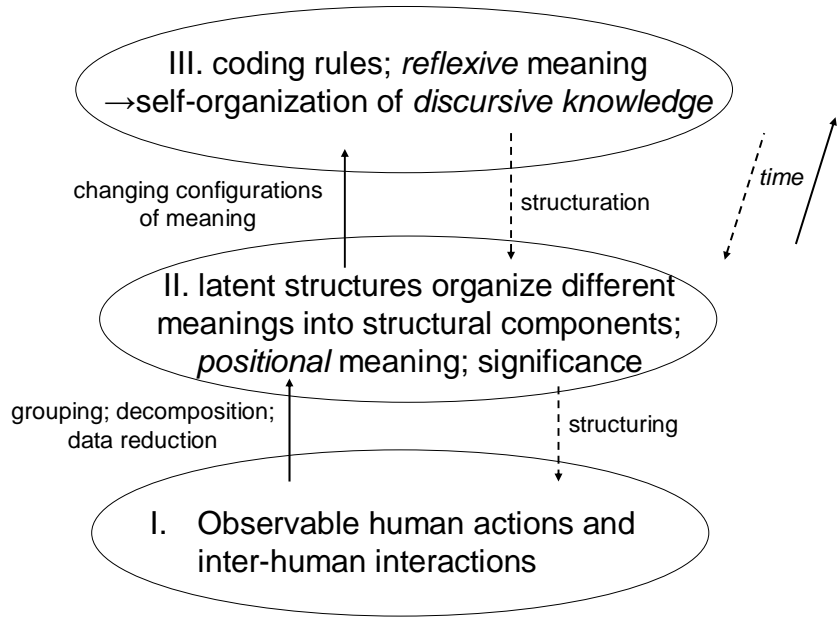

**Figure 4**: A layered process of codification of information by the processing of meaning, and the codification of meaning in terms of discursive knowledge. (Adapted from Leydesdorff (2010a), at p. 405.)

Figure 4 summarizes my argument hitherto. At the lowest level (I), one can use the results of measurements. For example, one may use data from questionnaires about what a stimulus means to respondents, or—closer to our example—relations among articles such as citations, co-authorship relations, or shared co-occurrences of words. The distributions of observable relations contain expected information about the social and intellectual structures operating upon this data. Organization reduces uncertainty in the data at each moment of time; self-organization generates an overlay of expectations that can feed back on the present and thus potentially reduce uncertainty within the system.

The structuring of the information processing is provided by positioning the information as in a factor model (level II). Meaning is generated in the recursive relations among information contents. Structuration is provided in terms of (next-order) codes of communication (level III) which can be used to relate different meanings. Whereas the positioning of the information at level II takes place at each moment of time, structuration (at level III) is based on the development of the organizing structure over time. The operations over time can be structurating because the relevant data is structured at each moment of time. Structures are reproduced and modified along trajectories, and restructuring can be reinforced or counteracted upon at a next-order regime level.

**The algorithmic model of meaning processing**

I noted above that meanings incur on events. In other words, events are understood from the perspective of hindsight, but with reference to possible future events. The time axis is thus a crucial dimension. It is often pictured as an arrow flying from the past via the present to the future, but the retrospective perspective of hindsight also operates in the processing of meaning, and thus time can be considered as yet another dimension or a degree of freedom. A model can provide us with a prediction about future states because it remains in a *res cogitans* in which future moments of time can be simulated in the present.

The mathematical biologists Rosen (1985) defined as "anticipatory" a system that is able to entertain a model of itself. The model can provide the system with one or more representations of future states in the present. These representations can be used for the active reconstruction of the system. Dubois (2003, at pp. 112f.) further distinguished between weakly and strongly anticipatory systems: a system that is weakly anticipatory is able to use its predicted states at future times for adaptation or intervention, while a system which is strongly anticipatory can use its anticipated states at time $t + 1$, $t + 2$, etc., for its present reconstruction.

Within this category of strongly anticipatory systems, Dubois (1998) further distinguished between incursive and hyper-incursive systems. Incursive systems use both their historical states and present or future states for their reconstruction, whereas hyper-incursive systems operate exclusively on the basis of expectations. In Leydesdorff (2009, 2010b), I described the *cogitans* as an embodied system that uses also historical states for the computation of a next one (in the present), whereas the social system or Husserl's *cogitatum* uses only future states, that is, expectations and their organization in systems of expectation. Such hyper-incursive cybernetics would operate against the axis of time and thus reduce uncertainty. Note that this is not yet a system, but a mechanism which requires anchoring in historical time by other (incursive) mechanisms.

As noted, a *cogitatum* cannot be instantiated without a *cogitans* to instantiate it; the two systems are structurally coupled and intertwined by reflexive interpenetration. Thus, there is always historical production of uncertainty in the "life-world" involved, but this forward arrow is counteracted by a feedback arrow from the self-organization of the codes. The latter is an evolutionary mechanism that operates in history, but against the arrow of time because redundancies are generated.

A hyper-incursive system cannot exist and be observed in the *res extensa*. The definitions are analytical and should not be reified in an external world. In other words, these are relevant subdynamics for the specification of the dynamics of communication of meaning and knowledge. However, the domains to be studied from this perspective are very different from biological or even psychological ones (Giddens, 1979). For example, it could be shown (Leydesdorff & Franse, 2009)—using the logistic equation (which can be used for modeling processes of growth and decline in biology) and its equivalent formulation in the hyper-incursive domain—that the biological and sociological domains are separated at the value of four of the so-called bifurcation parameter, and that this separatrix can only be crossed by invoking an incursive routine (that is, a psychological *cogitans* or human agency).

This is not the place to repeat the derivation of the various equations (Leydesdorff, 2009, 2010b). My crucial point is that the three cybernetic mechanisms specified by Luhmann can be operationalized as *different* mechanisms of incursion and hyper-incursion. When the various equations are solved, the conclusions are the following:

*Interaction*—hyper-incursively modeled as the interaction of two mutually expected selection mechanisms: $x_t = b\ (1 - x_{t+1})(1 - x_{t+1})$—leads to turn taking in the communication of meaning and thereby variation from the perspective of the social communication system. By extending this model parsimoniously only with a single (third) selection mechanism, one obtains two options which model *self-organization* and *organization* of meaning, respectively. When three anticipatory sources of variance can operate selectively upon one another, this might be modeled as follows:

$$x_t = c\ (1 - x_{t+1})(1 - x_{t+1})(1 - x_{t+1}) \tag{1}$$

Organization instantiates the interfaces historically, and one can analogously model this as follows:

$$x_t = d\ (1 - x_{t+1})(1 - x_{t+1})(1 - x_t) \tag{2}$$

The difference between the organization of meaning (Equation 2) and its self-organization (Equation 1) is provided in the third term: is the processing bent back to the present, or does this term in the model remain a reference to a future state? Organization can reduce uncertainty by instantiation in the present.

Equation 1 (modeling *self-organization*) has two imaginary roots and one real root. The real root can be considered as a constant operating in the coding. Equation 2 models a system (*organization* of meaning) that can be expected to perish after a finite number of historical instantiations. In other words, organizations of meaning emerge and disappear historically. Long-term stability is provided to the system meta-historically by the evolutionary mechanism of self-organization in the communication of meaning. The third mechanism—interaction—provides variation. Organization and self-organization are coupled to each other as Equations 1 and 2, but the results of these two mechanisms operating provide different solutions.

**Conclusion: society as *not* a system**

The metaphors of systems theory have been tainted by the biological discourse from which they emerged. Unlike biological systems, however, a social system cannot exist; the various *cybernetics* reconstruct orders of expectations that can be accessed reflexively. In an email conversation, Krippendorff (June 9, 2010; at

https://hermes.gwu.edu/cgi-bin/wa?A2=ind1006&L=cybcom&F=&S=&P=1565) suggested to distinguish between systems theory as a meta-biological generalization and the specification of cybernetic mechanisms for explaining social order and communication in language.

For example, the metaphor of "self-organization" suffers from the biological origins of this theorizing (Maturana & Varela, 1980). The "self" needs quotation marks because an order of expectations cannot be expected to contain an identifiable self; it remains an order of distributions that operate on one another. Luhmann ([1984], 1995), for example, proposed distinguishing among the social, the temporal, and the substantive as three dimensions in which these distributions can be extended (Latour, 2002). The uncertainties contained in the distributions can over time be considered as expectations. Thus, the modeling of the communication of meaning and information is deeply entrenched.

My insistence on specifying expectations in the social sciences is not meant to imply that sociology is an impossible—since non-empirical—science about non-observables. Uncertainties can be hypothesized, and we explored the various mechanisms that can be specified with reference to expectations. The specification of expectations enriches the research design beyond a behaviouristic focus on agency and institutions. Understanding and interpretation of language and knowledge require reflexive communication.

These cybernetic mechanisms need to be specified in a mathematical language that is yet content-free such as the language of information theory and cybernetics, since otherwise one is easily deluded by the biological or sometimes psychological connotations of the metaphors. One is in need of fresh metaphors, such as Giddens' (1979) notion of "structuration" (Leydesdorff, 2010b). Sociology becomes fundamentally different from biology and psychology as soon as one begins to focus on the communication of meaning and knowledge. In this context, one should remember Luhmann's dictum that society is not made up of human beings, but constructed in terms of their communications (Luhmann, 1996; cf. Marx [1858], 1973, at p. 265).

Unfortunately, in some of his later writing Luhmann (e.g., 1997) departed from these sociological assumptions and succumbed to the attraction of developing a general theory of observation. This is countrary to the sociological project, in my opinion, because a meta-biological assumption is introduced (Habermas, 1987; Leydesdorff, 2006b and 2010c). Following these later writings of Luhmnann, one would leave the deeply humanistic appeal which transpires from Husserl's philosophy in favour of considering the social as an automaton in which the Greek gods operate in the disguise of "performative media of communication."

I have wished to argue that symbolically generalized media of communication codify our expectations and thereby empower our performances reflexively in terms of handling complexity in terms of expectations. Performance, however, remains an attribute of human agency. The reflexive understanding of horizons of meaning made possible by communication provides us with access to a social reality in which knowledge-based anticipations play an increasing role. The (self-)organization of meaning at the above-individual level no longer "structurates" only our *actions* (Giddens, 1979, 1984), but more importantly our *expectations*.